\documentstyle[twocolumn,floats,epsfig,ifthen,aps,prb]{revtex}

\begin{document}
\draft
\title{Analysis of integrated single-electron memory operation} 
\author{Alexander N. Korotkov} 
\address{Department of Electrical Engineering,  
University of California, Riverside, CA 92521} 

\date{\today}

\maketitle

\begin{abstract}
        Various aspects of single-electron memory are discussed.
In particular, we analyze the single-electron charging by  
Fowler-Nordheim tunneling, propose the idea of background charge 
compensation, and discuss the defect-tolerant architecture based
on nanofuses. 
\end{abstract}



\narrowtext

        It is a common wisdom now that because of the size reduction 
of the components of integrated digital devices, the effects of charge 
discreteness will eventually become important. And there is a strong belief 
that the correlated tunneling of single electrons \cite{Av-Likh,SCT} 
in such ultradense devices can provide the physical basis for the new 
principle of their operation (see, e.g., Refs.\ \cite{Likh-rev,Kor-rev}). 

        There are two main possible areas of prospective digital
single-electronics: logic and memory devices. The theoretical analysis
shows \cite{Likh-rev,Kor-rev} that the single-electron memory is much
easier for the implementation than the logic. The basic reason is that
a logic device is necessarily a complex system consisting of many gates
interacting in a specific way, while memory cells are independent, each
of them being a simple circuit. The operation of the logic requires some
kind of voltage amplification (which can be also done parametrically
\cite{parametron}) to pass information from gate to gate.
In contrast, in the memory the storage of information can be done in a
passive way, and for the readout only some sensing of the storage contents
is sufficient (the amplification can be done at the next stage, common for
many memory cells). 
        
        As an example, the single-electron transistor (SET) can amplify
the voltage only at temperatures\cite{Kor-logic} $T<0.052 \, e^2/C_\Sigma$ 
where $C_\Sigma$ is the total capacitance of the SET central island  
while in the sensing mode it can be used at temperatures up to 
$\sim 0.25 \, e^2/C_\Sigma$ (the modulation amplitude
of the SET is still more than 10\% at this temperature). The possibility
to use significantly higher temperatures is very important for 
single-electronics. In addition, the problem of random background charge 
can be solved for memory  (while for the logic no reasonable 
solutions have been proposed so far) that is also very important for 
integrated circuits. The basic idea of the background-charge-insensitive
operation proposed in Ref.\ \cite{q0-ind} is to use oscillating output 
of the SET as a response to the ramp-up input signal so that the phase 
of the oscillations (which can be unpredictably shifted by the background 
charge) is not important. The input signal is generated during destructive 
readout of logical ``1'', which erases the few-electron charge stored 
at the floating gate close to the SET, while there is no signal if 
there was no stored charge (logical ``0''). The low-temperature prototype
of the background-charge-insensitive single-electron memory has been
demonstrated experimentally.\cite{NEC-mem} 

        There have also been a considerable number of experiments 
(including room-temperature experiments)   
on single-electron memory using different ideas (see, e.g., Refs.\
\cite{Yano,Dresselhaus,Nakazato-mem,Fujiwara,Tiwari,Chou-mem,%
Nakajima,Durrani,Lotkhov,Dutta,Sunamura,Kim,Yoo,Matsumoto,Kapetanakis}).
While most of them would have principal difficulties at the realistic 
level of integration, the experimental success supports the optimistic 
prospect of technologically practical single-electron memory. 

        In this paper we discuss various single-electron effects 
in the memory devices. We consider the DRAM-like and nonvolatile 
single-electron memories based on the charge storage at the floating gate.
(The SRAM-like single-electron memory \cite{Kor-logic} suffers from 
the same difficulties as the single-electron logic.) In particular,
we discuss the interplay between Coulomb blockade and Fowler-Nordheim
tunneling, propose the idea of background charge compensation, 
and propose the defect-tolerant architecture for single-electron memory
based on nanofuses and nanoshorts.

        Let us assume that the digital information is stored in the form 
of an electric charge. Then there are at least three aspects, for which
single electrons can be important. First, the digital bit can be 
represented by few or even only one electron. Second, the Coulomb 
blockade can be used as a mechanism providing bistability of the
memory cell. Third, the readout can be done using a single-electron 
principle, for example, the stored charge can be sensed by the SET.
Let us consider these aspects in more detail.

        The bit representation by a single electron provides the lowest
energy dissipation for write/erase operations and can in principle 
be performed even when the ``floating gate'' has the size as small as
one atom. Also, single electron storage is the physical limit of electronics,
and in this respect it is technologically and psychologically important. 
Notice, however, that in this case the bit information can be 
instantaneously destroyed by only one leakage event. As a consequence, 
the probability of error is suppressed only linearly with the decrease of
the time before the readout (in contrast to almost exponential suppression 
for continuous leakage). This also leads to impossibility of the information
refreshing using simple read-write back procedure, since there is no 
bit ``aging''. 

The straightforward way  to  improve   reliability  is  to  use 
redundancy;  
for example, to store the bit simultaneously in three memory cells
and use the majority principle at readout. However, obviously 
it is simpler to use redundancy inside the memory cell, which is to represent 
bit by three stored electrons, so that the leakage of one electron is 
allowed. This decreases the leakage error probability 
down to $P^2$, where $P$ is the error probability for the one-electron cell, 
and makes possible the information refreshing (since $P^2$ scales as a square
of retention time). 
Further increase of the number of stored electrons leads to further 
reduction of the error probability. One can expect that the optimal
number of stored electrons is somewhere between 5 and 30. Notice,
however, that this number should be controlled with single-electron 
accuracy: the usual Poisson distribution $n\pm \sqrt{n}$ is unacceptably 
wide when $n \lesssim 30$. 

        A different method to improve the reliability of a memory cell with 
one electron per bit is to readout simultaneously a block of cells 
and use the idea of ``control sums''. For example, using $2N$ extra cells
as control sums for columns and rows of a $N\times N$ block of cells, 
one can easily restore the loss 
of one bit and, hence, reduce the error probability down to $\sim P^2$. 
Additional control sums can restore the loss of more than one bit, 
that suppresses further the error probability.
(Actually, the use of columns, rows, or diagonals of a block for control
sums is obviously not the best way of introducing redundancy. Using 
standard coding algorithms \cite{coding} one can restore up to 11\% of errors
by doubling the numbers of memory cells.) 
 So, for the single-electron 
representation of a bit, the reliable information storage can also be 
achieved, however, the few-electron representation of a bit (inside one memory
cell) makes it significantly simpler and seems more natural for random 
access memory. 

        The charge storage requires very low leakage rate for both 
logical states while write/erase time should be sufficiently fast. 
In the present-day DRAMs this is achieved by the use of field-effect
transistor (FET) as a switch.
Unfortunately, the SET cannot replace FET for this purpose because of
significant cotunneling rate,\cite{Av-Likh} so other principles are
necessary. A promising principle is the control of the tunnel barrier 
by the gate electrode.\cite{Tucker} For single-electron memory a similar 
principle has been used in the experiments or Refs.\ 
\cite{Nakazato-mem,Durrani} (which have been discussed using the 
terminology of multiple-tunnel-junction SET). 

        Besides the gated operation of charge write/erase procedure,
one can consider non-gated operation similar to that used in the
conventional nonvolatile memories.\cite{flash} The idea is to use a 
threshold-like
behavior of the charging rate as a function of the voltage between the
storage floating gate and the word (or bit) line. 
Then the long retention time is achieved if the voltage due to 
charging of floating gate capacitance as well as the half-select 
write/erase voltage are below the threshold, while fast write/erase 
operation occurs when the external full-select voltage exceeds the threshold.

        A natural for single-electron memory idea is to use the Coulomb
blockade for such a threshold. For example, it can be provided by 
the array of small-capacitance tunnel junctions similar to that 
used in single-electron traps (see, e.g. Ref.\ \cite{Dresselhaus} and 
references therein). Unfortunately, the random background charges 
lead to unacceptably wide distribution of the Coulomb blockade thresholds 
and Coulomb blockade barrier heights.\cite{Likh-rev} 

        Another way is to use the threshold-like dependence of the 
conventional Fowler-Nordheim tunneling. The problem is that in this
case there is no sharp threshold, and as a consequence the ratio of
the retention and write/erase times is not sufficiently large. The 
situation can be significantly improved by the use of the tunnel barriers
of crested shape \cite{q0-ind,Likh-crested} so that not only the width
but also the height of the barrier decreases with the applied voltage.
        The crested barrier can be fabricated using $\delta$-doping,
composition grading, or layered structure. 

\begin{figure}
\vspace{0.2cm}
\centerline{
\epsfxsize=3.2in
\hspace{-0.1cm}
\epsfbox{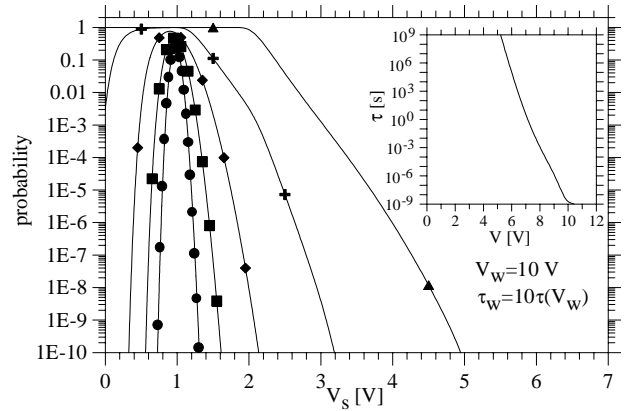} 
}
\vspace{0.4cm}
\caption{Inset: the characteristic time  $\tau =C_sV/I$ of 
	the continuous charging of a floating gate through 
	4nm/5nm/4nm trilayer barrier 
	$n^+$Si/Si$_3$N$_4$/AlN/Si/Si$_3$N$_4$/$n^+$Si. 
	Main figure: the probability distribution of the floating gate 
	potential $V_s$ after the charging, for several values of 
	the voltage $e/C_s$ corresponding to a single electron at the 
	floating gate: $e/C_s=$0.03 V (dots), 
	0.1 V (squares), 0.3 V (diamonds), 1 V (crosses), 
	and 3 V (triangles). Symbols correspond to the floating gate 
	background charge $q_0=-e/2$.} 
\label{fig1}\end{figure}

        We have studied the single-electron charging of the floating 
gate by Fowler-Nordheim tunneling using the following simple ``orthodox'' 
model.\cite{Av-Likh} The tunneling rate $\Gamma =I(V_{eff})/e$ is determined 
by the effective voltage 
        \begin{equation} 
V_{eff}=V_w - \frac{e}{2C_s} - \frac{q_0-ne}{C_s} 
        \end{equation}
where $V_w$ is the contribution due to externally applied write/erase 
voltage, 
$n$ is the number of extra electrons on the floating gate which changes 
during charging, $q_0$ is the background charge, and $C_s$ is the storage 
floating gate capacitance. For the ``seed'' {\it I-V} curve $I(V)$ we used 
the model of Ref.\ \cite{cooling} applied to the 4nm/5nm/4nm trilayer crested 
barrier with parameters corresponding to 
$n^+$Si/Si$_3$N$_4$/AlN/Si/Si$_3$N$_4$/$n^+$Si. The inset in Fig.\ 
\ref{fig1} shows the classical recharging time (defined as $\tau = C_s V/I$) 
as a function of the voltage $V$ across the barrier.

        Since the tunneling is a stochastic process, the number $n$ of
stored electrons after the application of voltage $V_w$ during time $\tau_w$,
is random. So, after the external voltage is removed, the potential 
$V_s=(q_0-ne)/C_s$ of the floating gate can be characterized by the
probability distribution. Fig.\ \ref{fig1} shows numerically
calculated distribution of $V_s$ for $V_w=10$ V, $\tau_w=10 \times
 \tau (V_w)$,  
and several values of the single-electron voltage $e/C_s$: 0.03 V (dots), 
0.1 V (squares), 0.3 V (diamonds), 1 V (crosses), and 3 V (triangles). 
The symbols correspond to the 
background charge $q_0=-e/2$ while the lines show the probability densities 
after the averaging over random $q_0$ (normalized in a way that the lines
go through the symbols). 

        One can see that the probability distribution is relatively narrow
when $e/C_s$ is small, and in this case there is essentially no difference 
between the random and well-defined $q_0$. The distribution width grows
with $e/C_s$, and in the case $e/C_s \sim V_s$ the fluctuations 
are unacceptably strong for the reliable information storage. The curve 
width continue to grow when $e/C_s$ becomes larger than typical $V_s$ 
of the classical charging. However, if $q_0$ is well-defined, the probability
distribution collapses into single well-predictable value of $V_s$.
Moreover, the voltage $V_s$ in this case can be considerably larger 
than for small $e/C_s$, that can be useful for the readout. 
It is important that the effective voltage which determines the 
retention time is $V_s-e/2C_s$ (less than $V_s$); so for $q_0=-e/2$, 
large $e/C_s$, and one stored electron (the upper triangle  
in Fig.\ \ref{fig1}) this voltage is exactly zero that can significantly  
improve the charge retention. 
        The probability of a ``dynamic'' error due to finite write/erase  
time $\tau_w$ in 
the symmetric one-electron case ($q_0=-e/2$) decreases exponentially 
with $\tau_w$,  
$P_e = \exp (-\Gamma \tau_w )$ where $\Gamma^{-1} = \tau (V_w)e/C_sV_w$ 
is determined by the full write voltage $V_w$. This is obviously an
advantage \cite{Tiwari} in comparison with the case of small $e/C_s$, 
in which the voltage gradually decreases in the process of charging, 
gradually decreasing the charging rate. 

        The analysis above shows that there are two preferable modes 
of operation. Either the bit should be represented by many ($\gtrsim$ 10) 
electrons (then there is no need to control $q_0$) or it should be 
just one electron (and $q_0$ is well-controlled), while in the few-electron 
regime the fluctuations of the stored charge are very strong. 
As one can see from Fig.\ \ref{fig1}, the typical voltage $e/C_s$ in 
the one-electron regime should be about few volts that obviously suggests
the use of single atoms as ``floating gates''
(the randomness of the location can be avoided using self-assembly). 
For single atoms the fluctuations of $q_0$ are naturally small
since the chemical environment is well defined. (Actually, for single
atoms the ``orthodox'' theory should be modified \cite{Av-Kor-Likh}, 
and instead of $q_0$ fluctuations we need to consider the variation
of the electron affinity.) 
Using the language more natural for atoms, the best case $q_0=-e/2$ 
corresponds to the impurity energy exactly at the Fermi level,
so that both occupied and empty states are equally stable. Notice
that the use of single atoms can be easily combined with the idea
of few-electron bit representation if few atoms (located sufficiently
far from each other) per memory cell are used. 

        Now let us discuss the readout of the stored charge. There is
an experimental evidence \cite{Chou-mem} supported by the theoretical
analysis \cite{Naveh} that FET can be used for sensing the charge 
at the size scale down to 10 nm. Another option (which 
seems to be preferable only at the size scale below 10 nm) is the 
use of SET. Besides the problem of low operation temperature
which will become less severe when the few-nm technology is available,  
another principal problem for integrated SET devices 
is the background charge fluctuations.\cite{Likh-rev,Kor-rev} 
One proposed solution \cite{q0-ind} is the background-charge-insensitive
operation. Here we propose a different solution.

\begin{figure}
\vspace{0.2cm}
\centerline{
\epsfxsize=2.8in
\hspace{-0.1cm}
\epsfbox{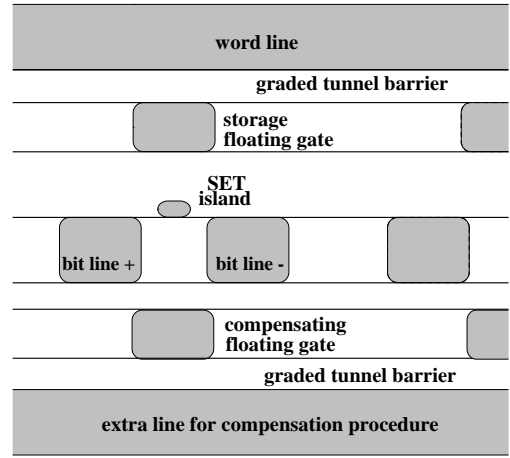}
}
\vspace{0.4cm}
\caption{Single-electron memory cell with the SET readout using the
background charge compensation. The information is stored as a charge
of the upper floating gate and is read out by the SET consisting of two
tunnel junctions between the SET island and two bit lines. The SET background
charge is compensated by the charge of lower floating gate. Graded tunnel
barriers improve the Fowler-Nordheim charging of floating gates from two
word lines.} 
\label{fig2}\end{figure}

        The idea is to tune (compensate) individually the background 
charge $Q_0$ of each SET. From the first sight this seems impossible in an  
integrated circuit. However, the simple uniform architecture 
of the memory allows the local self-compensation procedure. 
Instead of tuning the background charge by the voltage applied to an 
extra gate (that is typically used in present-day single-electron
experiments) we propose to use extra floating gate.  Fig.\ \ref{fig2}
shows the memory cell consisting of storage floating gate (it can
be replaced by single atoms as discussed above), SET to sense 
its charge, and $Q_0$-compensating floating gate (CFG). If the dimensionless
coupling between the CFG and the central island of the SET is about 0.1,
then placing the proper number of electrons on CFG we can control $Q_0$
with the accuracy of $0.1\, e$ that is sufficient for predictable
readout from the SET.

        The CFG is charged by the Fowler-Nordheim tunneling from
the extra word line (see Fig.\ \ref{fig2}) and the amount of charge
is determined by the voltage applied between this line and bit lines
connected to SET. The compensation procedure can be done in the following
way. The storage floating gate is prepared in logical state ``0'', 
the SET is biased, 
and the ramp-up voltage is applied to one of the word lines (or both).
The SET output performs the oscillations, the phase of which carries the 
information about $Q_0$. After amplification by the sense amplifier connected
to the bit line, this signal is used to determine the proper magnitude
of the voltage pulse to be applied between compensating word line and 
bit lines (an iterative sequence of trials and tests can be useful).
The compensating pulse should be applied to the bit lines to allow
simultaneous compensation procedure for all cells connected to the
same word line. 
Since the compensation procedure requires the charge erasure from the
storage gate, it can naturally be combined with the information refreshing.

        Besides the presence of the CFG and extra word line, the layout and 
the basic parameters of the 
$Q_0$-compensated single-electron memory is similar to that of
background-charge-insensitive memory of Ref.\ \cite{q0-ind}. 
In particular, the room temperature operation can also be achieved 
at $\sim$4 nm minimum feature size (20 nm $\times$ 40 nm total area 
per cell). The important advantage in comparison with the 
background-charge-insensitive memory is the possibility of nondestructive 
readout. 
This also implies the reduction of the stored charge and/or coupling
between the SET and storage gate, because this charge reduced to the 
SET input should no longer correspond to several Coulomb oscillations
but only to a fraction of the period instead.

        If the SET operates in the high-temperature ``analog'' regime 
($T\sim 0.2e^2/C_\Sigma$) 
then the destructive readout remains the only reasonable option.
The SET should be tuned to the most sensitive operation point of the control
characteristic and the output SET current before and after the attempted 
bit erasure
should be compared. It is important that other SETs connected to the same
bit line are also biased and contribute to the current noise. The maximum 
number $M$ of SETs per sense amplifier is determined by the bandwidth and
acceptable signal to noise ratio, and in this case is comparable to that of 
background-charge-insensitive memory, i.e.\ $M \sim 10^2$. 
In the low-temperature regime ($T \lesssim 0.05 e^2/C_\Sigma$) the current
and noise from half-selected SETs (biased but not selected by proper gate 
voltage) can be strongly suppressed by the Coulomb blockade,  
so the SET can essentially operate as a switch. This provides much 
better signal to noise ratio and allows for a nondestructive readout
as well as a significant increase of $M$ in $Q_0$-compensated single-electron
memory.
 
        The architecture of the single-electron memory should obviously
differ from that of conventional memory. First, if the SET is used 
for the readout, then the relatively high output impedance of the SET 
requires quite short local bit lines in order to reduce their charging time. 
Second, with the strong decrease of the feature size and possible use
of the molecular electronics technology, one can expect the reduction
of the yield per memory cell, \cite{Heath} so the architecture should
be able to tolerate significant fraction of defects. We propose here a novel 
defect-tolerant memory architecture based on nanofuses and nanoshorts
(which can be useful for any ultradense memory). 

        The main idea is the local physical rerouting of the bit (or word) 
lines in order to avoid defective cells. In contrast to the logic,
the uniform memory organization allows us to test each memory cell 
(of course, such testing can be done in parallel for many cells). 
For significant fraction of defects, any long bit line would contain
defects. So only rejection of relatively small pieces of the
line is possible and the architecture should necessarily be local.
The significant fraction of defects also makes it impossible to store
the information about the defective cells and then switch the addresses 
logically, since it could require storage space comparable to the total 
available memory. The problem can be solved by physical rerouting of
bit lines to replace the relatively short defective pieces by good ones 
from the local reserves. This is done once during the testing procedure
while from the outside (global level of hierarchy) all the lines 
look nondefective.

\begin{figure}
\vspace{0.2cm}
\centerline{
\epsfxsize=3.2in
\hspace{-0.1cm}
\epsfbox{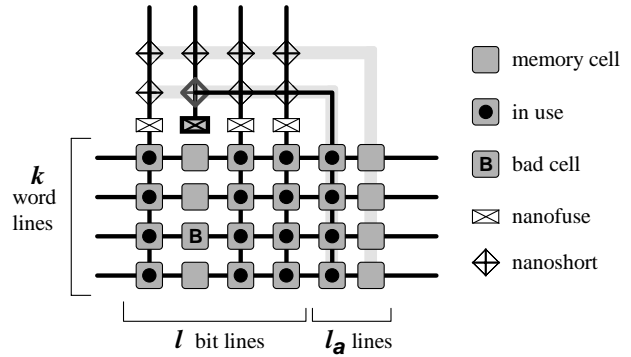}
}
\vspace{0.4cm}
\caption{The basic idea of local defect-tolerant architecture for 
	single-electron memory using nanofuses and nanoshorts. The short 
	pieces of bit lines are rerouted by one-time switching 
	of nanofuse-nanoshort pairs to avoid defective cells, so that the
	lines look nondefective from outside. 
	} 
\label{fig3}\end{figure}

        Fig.\ \ref{fig3} illustrates the idea of rerouting. 
The cells are organized in blocks of the relatively small size 
$k\times (l+l_a)$, 
where $k\times l$ cells are normally used while
$k \times l_a$ cells are ``in reserve''. If during the testing 
a cell is found to be defective, the corresponding line of $k$ cells should 
be replaced by the line from reserve. For this purpose we need 
two types of switches: nanofuses (which can be blown to disconnect 
the line) and nanoshorts (which can connect lines). Notice that
each switch is used at most once, so it can be a quite simple and therefore
reliable nanoscale device. The importance of nanoswitches for any type of
nanoelectronics has been previously emphasized in Ref.\ \cite{Heath} 
and the experimental progress has been already reported.\cite{Collier} 

        The blown nanofuse disconnects the defective line of $k$ cells; 
instead, the switched nanoshort connects another line of $k$ cells 
to the same global bit line so that from outside the addressing does 
not change. Since the global bit line should pass through much 
more than $k$ cells, we either need to return to the main bit line 
after the detour, switching another nanofuse/nanoshort pair, or use  
the idea of local $k$-long branches connected to the global bit line 
``in parallel''. 

        The optimal values of $k$, $l$, and $l_a$ obviously 
depend on the expected fraction $p$ of defective memory cells. 
Notice that the idea works even if $p$ is comparable to unity. 
(In this case one can use $k=1$, $l=1$, and sufficiently large 
$l_a \sim -10\ln p$.)  
        If in a bad luck case $l_a$ lines prepared for the replacement 
are not sufficient for defect-free operation, the rerouting of
a longer piece of the bit line can be used at the next level
of hierarchy. 

        The switching of nanofuses and nanoshorts requires extra 
wires for their selection, which are not discussed here. Also, we assumed 
perfect wires and nanoswitches. This assumption seems to be reasonable, since
the memory cell is a more complex device and, hence, has much larger 
chance to be defective. The rare defects of wires, nanofuses, and nanoshorts 
can be treated by conventional means.

        In conclusion, we have discussed various aspects of the 
single-electron memory operation. 
   First, the analysis of the Fowler-Nordheim charging of a floating gate 
in a single-electron regime shows that the most preferable mode of operation
is when two logical states differ by exactly one electron on a floating gate,
which itself has the background charge $q_0=-e/2$. This suggests the use of 
single impurity atoms at the Fermi level as floating gates.  
Second, the operation of the single-electron transistors for memory readout 
can be significantly improved by using the proposed $Q_0$-compensation 
procedure employable in integrated design. Third, the proposed defect-tolerant
memory architecture based on nanofuses and nanoshorts can provide reliable
operation even in the case of significant fraction of defective memory cells.
Leaving aside the major present-day obstacle for single-electron memory, which
is the need for reliable few-nm technology, the overall prospect  
of the development of integrated single-electron memory within next
10-20 years seems to be quite optimistic.

The author thanks K. K. Likharev for numerous fruitful discussions. 
  The work was supported by the Semiconductor Research Corporation Contract 
No.\ 2000-NJ-746.

\end{document}